# A Hybrid Clustering Algorithm for Data Mining

Ravindra Jain

School of Computer Science & IT, Devi Ahilya Vishwavidyalaya, Indore, India
ravindra.jain57@gmail.com

*ABSTRACT*

*Data clustering is a process of arranging similar data into groups. A clustering algorithm partitions a data set into several groups such that the similarity within a group is better than among groups. In this paper a hybrid clustering algorithm based on K-mean and K-harmonic mean (KHM) is described. The proposed algorithm is tested on five different datasets. The research is focused on fast and accurate clustering. Its performance is compared with the traditional K-means & KHM algorithm. The result obtained from proposed hybrid algorithm is much better than the traditional K-mean & KHM algorithm.*

## 1. INTRODUCTION

The field of data mining and knowledge discovery is emerging as a new, fundamental research area with important applications to science, engineering, medicine, business, and education. Data mining attempts to formulate, analyze and implement basic induction processes that facilitate the extraction of meaningful information and knowledge from unstructured data [1].

Size of databases in scientific and commercial applications is huge where the number of records in a dataset can vary from some thousands to thousands of millions [2]. The clustering is a class of data mining task in which algorithms are applied to discover interesting data distributions in the underlying data space. The formation of clusters is based on the principle of maximizing similarity between patterns belonging to distinct clusters. Similarity or proximity is usually defined as distance function on pairs of patterns and based on the values of the features of these patterns [3] [4]. Many variety of clustering algorithms have been emerged recently. Different starting points and criteria usually lead to different taxonomies of clustering algorithms [5]. At present commonly used partitional clustering methods are K-Means (KM) [6], K-Harmonic Means (KHM) [7], Fuzzy C-Means (FCM) [8][ 9], Spectral Clustering (SPC) [10][ 11] and several other methods which are proposed on the base of above mentioned methods[12]. The KM algorithm is a popular partitional algorithm. It is an iterative hill climbing algorithm and the solution obtained depends on the initial clustering. Although the K-means algorithm had been successfully applied to many practical clustering problems, it has been shown that the algorithm may fail to converge to a global minimum under certain conditions [13].

The main problem of K-means algorithm is that, the algorithm tends to perform worse when the data are organized in more complex and unknown shapes [14]. Similarly KHM is a novel center-based clustering algorithm which uses the harmonic averages of the distances from each data point to the centers as components to its performance function [7]. It is demonstrated that KHM is essentially insensitive to the initialization of the centers. In certain cases, KHM significantly improves the quality of clustering results, and it is also suitable for large scale data sets [5]. However existing clustering algorithm requires multiple iterations (scans) over the datasets to achieve a convergence [15], and most of them are sensitive to initial conditions, by example, the K-Means and ISO-Data algorithms [16][17]. This paper proposed a new hybrid algorithm which is based on K-Mean & K-Harmonic Mean approach.

The remainder of this paper is organized as follows. Section II reviews the K-Means algorithm and the K-Harmonic Means algorithm. In Section III, the proposed K-Means and K-Harmonic Means based hybrid clustering algorithm is presented. Section IV provides the simulation results for illustration, followed by concluding remarks in Section V.

## 2. ALGORITHM REVIEWS

In this section, brief introduction of K-Means and K-Harmonic Means clustering algorithms is presented to pave way for the proposed hybrid clustering algorithm.

### 2.1. *K-Means Algorithm*

The K-means algorithm is a distance-based clustering algorithm that partitions the data into a predetermined number of clusters. The K-means algorithm works only with numerical attributes. Distance-based algorithms rely on a distance metric (function) to measure the similarity between data points. The distance metric is either Euclidean, Cosine, or Fast Cosine distance. Data points are assigned to the nearest cluster according to the distance metric used. In figure 1 the K-Mean algorithm is given.

1. begin
2. initialize N, K, $C_1, C_2, \ldots, C_K$;
   where N is size of data set,
   K is number of clusters,
   $C_1, C_2, \ldots, C_K$ are cluster centers.
3. do assign the n data points to the closest $C_i$;
   recompute $C_1, C_2, \ldots, C_K$ using Simple Mean function;
   until no change in $C_1, C_2, \ldots, C_K$;
4. return $C_1, C_2, \ldots, C_K$;
5. End

Figure 1. K-means Algorithm

### 2.2. *K-Harmonic Means Algorithm*

K-Harmonic Means Algorithm is a center-based, iterative algorithm that refines the clusters defines by K centers [7]. KHM takes the harmonic averages of the squared distance from a data point to all centers as its performance function. The harmonic average of K numbers is defined as- the reciprocal of the arithmetic average of the reciprocals of the numbers in the set.

$$HA(\{a_i | i = 1, \ldots, K\}) = \frac{K}{\sum_{i=1}^{K} \frac{1}{a_i}} \quad (1)$$

The KHM algorithm starts with a set of initial positions of the centers, then distance is calculated by the function $d_{i,l} = \|x_i - m_l\|$, and then the new positions of the centers are calculated. This process is continued until the performance value stabilizes. Many experimental results show that KHM is essentially insensitive to the initialization. Even when the initialization is worse, it can also converge nicely, including the converge speed and clustering results. It is also suitable for the large scale data sets [5]. Because of so many advantages of KHM, it is used in proposed algorithm.

1. begin
2. initialize N, K, $C_1, C_2, \ldots, C_K$;
   where N is size of data set,
   K is number of clusters,
   $C_1, C_2, \ldots, C_K$ are cluster centers.
3. do assign the n data points to the closest $C_i$;
   recompute $C_1, C_2, \ldots, C_K$ using distance function;
   until no change in $C_1, C_2, \ldots, C_K$;
4. return $C_1, C_2, \ldots, C_K$;
5. End

Figure 2. K-Harmonic Mean Algorithm

## 3. PROPOSED ALGORITHM

Clustering is based on similarity. In clustering analysis it is compulsory to compute the similarity or distance. So when data is too large or data arranged in a scattered manner it is quite difficult to properly arrange them in a group. The main problem with mean based algorithm is that mean is highly affected by extreme values. To overcome this problem a new algorithm is proposed, which performs two methods to find mean instead of one.

The basic idea of the new hybrid clustering algorithm is based on applying two techniques to find mean one by one until or unless our goal is reached. The accuracy of result is much greater as compared to K-Mean & KHM algorithm. The main steps of this algorithm are as follows:

Firstly, choose K elements from dataset DN as single element cluster. This step follow the same strategy as k-mean follows for choosing the k initial points means choosing k random points. Figure 3 shows the algorithm.

In section 4, the result of experiments illustrate that the new algorithm has its own advantages, especially in the cluster forming. Since proposed algorithm apply two different mechanisms to find mean value of a cluster in a single dataset, therefore result obtained by proposed algorithm is benefitted by the advantages of both the techniques. For another aspect it also sort out the problem of choosing initial points as this paper mentioned earlier that Harmonic mean converge nicely, including result and speed even when the initialization is very poor. So in this way the proposed algorithm overcomes the problem occurred in K-Mean and K-Harmonic Mean algorithms.

1. begin
2. initialize Dataset $D_N$, K, $C_1, C_2, \ldots, C_K$, CurrentPass=1;
   where D is dataset, N is size of data set,
   K is number of clusters to be formed,
   $C_1, C_2, \ldots, C_K$ are cluster centers.
   CurrentPass is the total no. of scans over the dataset.
3. do assign the n data points to the closest $C_i$;
   if CurrentPass%2==0
       recompute $C_1, C_2, \ldots, C_K$ using Harmonic Mean function;
   else
       recompute $C_1, C_2, \ldots, C_K$ using Arithmetic Mean function;
   increase CurrentPass by one.
   until no change in $C_1, C_2, \ldots, C_K$;
4. return $C_1, C_2, \ldots, C_K$;
5. End

Figure 3. Proposed Hybrid Algorithm

## 4. EXPERIMENTATION AND ANALYSIS OF RESULTS

Extensive experiments are performed to test the proposed algorithm. These experiments are performed on a server with 2 GB RAM and 2.8 GHz Core2Duo Processor. Algorithms are written in JAVA language and datasets are kept in memory instead of database. Each dataset contains 15000 records, since heap memory of JAVA is very low so the experiments use only 10 percent of original dataset i.e. 1500 records to perform analysis.

Here, the comparison between new hybrid algorithm with traditional K-mean algorithm and KHM algorithm is shown. In each experiment 5 clusters are created and accuracy of clusters is tested with clusters formed using traditional K-mean algorithm and KHM algorithm.

Figure 4, 5, 6, 7, 8 and Table 1 shows that New Hybrid algorithm reduces the mean value of each cluster, which means elements of clusters, are more tightly bond with each other and resultant clusters are more dense. Because the mean is greatly reduced and efficiently calculated, the algorithms have many advantages in the aspects of computation time and iteration numbers, and the effectiveness is also greatly improved. From the experiments, it can also discover that the clustering results are much better.

Experiment 1:
In this experiment the maximum item value of our dataset is 997 and minimum item value is 1.

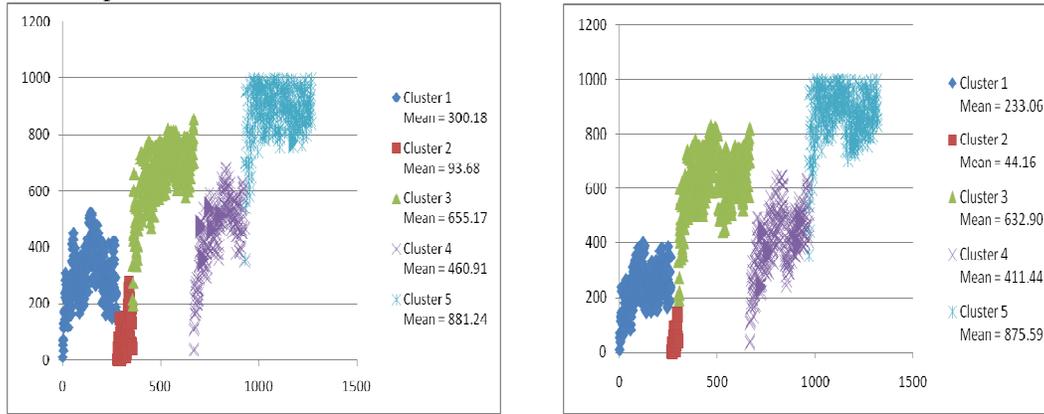

    (a)  K-Mean Algorithm                (b) Proposed Hybrid Algorithm
Figure 4.  Result obtained from applying K-Mean & Proposed algorithm on Dataset 1.

Experiment 2:
In this experiment the maximum item value of our dataset is 9977 and minimum item value is 6.

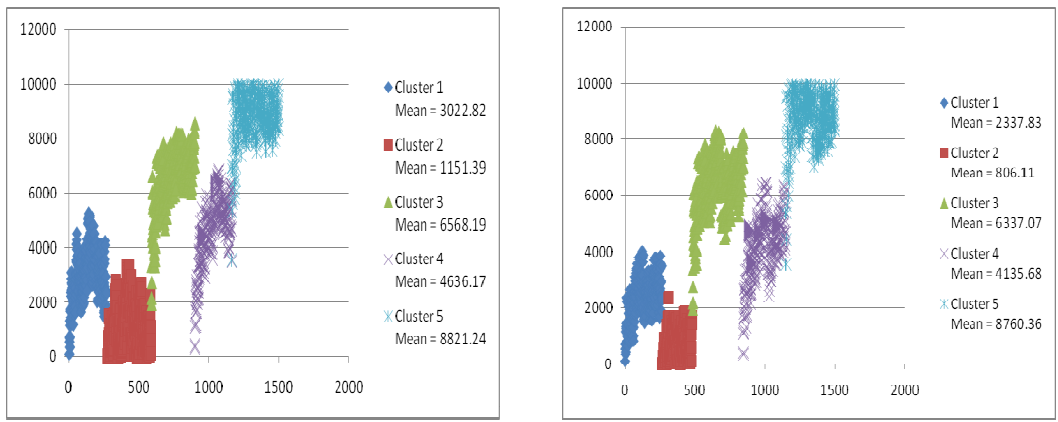

    (a)  K-Mean Algorithm                (b) Proposed Hybrid Algorithm
Figure 5.  Result obtained from applying K-Mean & Proposed algorithm on Dataset 2.

Experiment 3:
In this experiment the maximum item value of our dataset is 199 and minimum item value is 2.

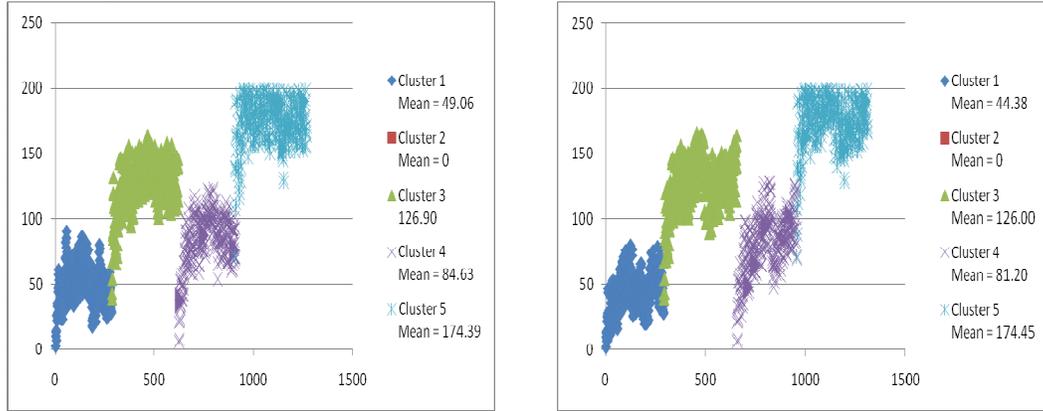

    (a) K-Mean Algorithm            (b) Proposed Hybrid Algorithm

Figure 6. Result obtained from applying K-Mean & Proposed algorithm on Dataset 3.

Experiment 4:
In this experiment the maximum item value of our dataset is 58785790 and minimum item value is 58720260.

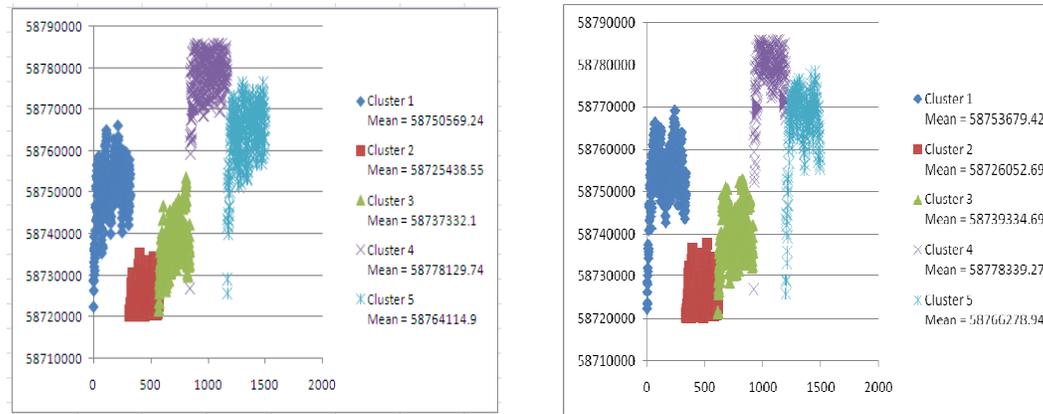

    (a) K-Mean Algorithm            (b) Proposed Hybrid Algorithm

Figure 7. Result obtained from applying K-Mean & Proposed algorithm on Dataset 4.

Experiment 5:
In this experiment the maximum item value of our dataset is 498 and minimum item value is 1.

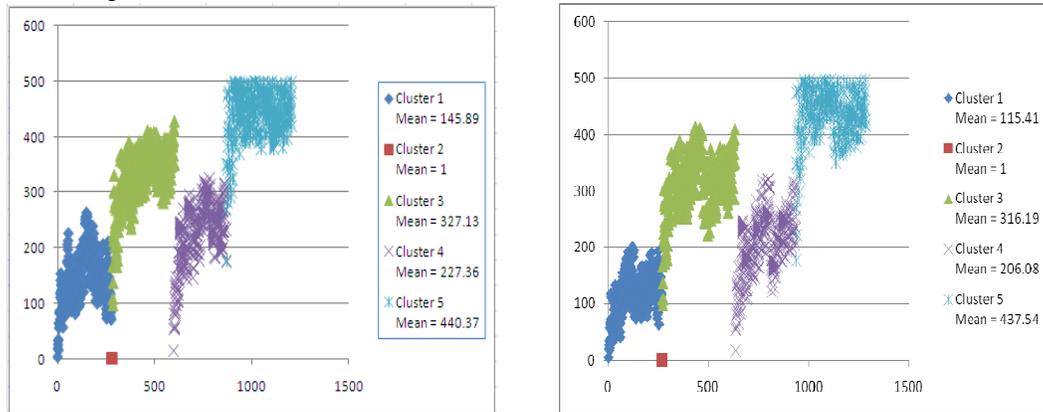

    (a) K-Mean Algorithm            (b) Proposed Hybrid Algorithm

Figure 8. Result obtained from applying K-Mean & Proposed algorithm on Dataset 5.

Table 1. Comparison between results obtained from applying KHM and proposed hybrid algorithm

| Exp. No. | Cluster No. | KHM Algorithm | | New Hybrid Algorithm | |
|---|---|---|---|---|---|
| | | *Total Element in Cluster* | *Mean Value of Cluster* | *Total Element in Cluster* | *Mean Value of Cluster* |
| 1. | 1 | 263 | 233.08 | 263 | 233.06 |
| | 2 | 37 | 44.17 | 37 | 44.16 |
| | 3 | 365 | 632.90 | 365 | 632.90 |
| | 4 | 303 | 411.49 | 302 | 411.44 |
| | 5 | 352 | 875.60 | 352 | 875.59 |
| 2. | 1 | 265 | 2337.92 | 263 | 2337.83 |
| | 2 | 220 | 806.15 | 219 | 806.11 |
| | 3 | 363 | 6337.10 | 363 | 6337.07 |
| | 4 | 304 | 4135.72 | 303 | 4135.68 |
| | 5 | 352 | 8760.48 | 352 | 8760.36 |
| 3. | 1 | 289 | 44.36 | 288 | 44.38 |
| | 2 | 0 | 0.00 | 0 | 0.00 |
| | 3 | 367 | 126.03 | 367 | 126.00 |
| | 4 | 301 | 81.20 | 301 | 81.20 |
| | 5 | 354 | 174.46 | 354 | 174.45 |
| 4. | 1 | 336 | 58753679.45 | 335 | 58753679.42 |
| | 2 | 275 | 58726052.77 | 275 | 58726052.69 |
| | 3 | 309 | 58739334.72 | 307 | 58739334.69 |
| | 4 | 283 | 58778339.25 | 284 | 58778339.27 |
| | 5 | 300 | 58766278.95 | 299 | 58766278.94 |
| 5. | 1 | 269 | 115.45 | 268 | 115.41 |
| | 2 | 2 | 1.00 | 2 | 1.00 |
| | 3 | 368 | 316.25 | 365 | 316.19 |
| | 4 | 301 | 206.12 | 301 | 206.08 |
| | 5 | 353 | 437.58 | 352 | 437.54 |

## 5. CONCLUSIONS

This paper presents a new hybrid clustering algorithm which is based on K-mean and KHM algorithm. From the results it is observed that new algorithm is efficient. Experiments are performed using different datasets. The performance of new algorithm does not depend on the size, scale and values in dataset. The new algorithm also has great advantages in error with real results and selecting initial points in almost every case.

Future enhancement will include the study of higher dimensional datasets and very large datasets for clustering. It is also planned to use of three mean techniques instead of two.

**Authors**

Mr. Ravindra Jain is working as a Project Fellow under UGC-SAP Scheme, DRS Phase – I at School of Computer Science & IT, Devi Ahilya University, Indore, India since December 2009. He received B.Sc. (Computer Science) and MCA (Master of Computer Applications) from Devi Ahilya University, Indore, India in 2005 and 2009 respectively. His research areas are Databases, Data Mining and Mobile Computing Technology. He is having 2+ years of experience in software development and research.

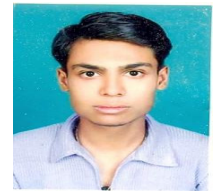